\documentclass[twocolumn,showpacs,aps,10pt]{revtex4-1}
\usepackage{color}
\usepackage{epsfig}
\usepackage{graphics}
\begin{document}

\title{
Augmentation of nucleon-nucleus scattering by information entropy.
}
\author{S.E. Massen}\email{massen@auth.gr} \author{V.P. Psonis}
\affiliation{Department of Theoretical Physics and HINP, Aristotle
University of Thessaloniki, GR 54124 Thessaloniki, Greece}
\author{H.V. von Geramb}
\affiliation{Department Physik, Universt\"at Hamburg, Luruper
Chaussee 149, D-22761 Hamburg, Germany}

\begin{abstract}
Quantum information entropy is calculated from the nucleon nucleus
forward scattering  amplitudes. Using a representative set of
nuclei, from $^4$He to $^{208}$Pb, and energies,  $T_{lab} <
1$\,[GeV], we establish a linear dependence of quantum information
entropy as functions of  logarithm nuclear mass $A$ and logarithm
projectile energy $T_{lab}$.

\end{abstract}


\maketitle

\section{Introduction}

Studies of many body systems, like the sophisticated fields of atomic,
nuclear and particle physics, have seen new and versatile
enrichment under the auspices of quantum
information theory. On one side we are aware of the implications
of a first principle treatment of  complex quantum mechanical
systems and on the other side we are  aware of less sophisticated
phenomenological solutions that simplify the
many body systems significantly. Phenomenology requires
only a few degrees of freedom and introduces
effective intensive and extensive  matter properties.

The growing awareness and hope to exploit
entanglement of quantum mechanical systems, as a new resource, has fostered
many fields to investigate their realm also
within the terminology of information theory. This is undoubtedly the case for
a wide range of molecular and atomic systems and less  for nuclear and particle
physics. At least currently, the latter systems are not
guided by practical applications but rather epistemological
virtues.

Nuclear geometries are realized with  nuclear structure
calculations or measurements in conjunction with direct nuclear
reactions  \cite{feshbach}.
Their values are now available in comprehensive tables  and
the data-groups continue their program as new facilities
advent. In terms of information theory, nuclear physics is
still in its infancy.

Shannon information entropy, together with  quantum information
theory,  has been the subject of many theoretical
studies  \cite{nielsen,bengtsson}.
For its realization we  know many cases with
a fundamental microscopic approach  as well as approaches which prefer
a  phenomenological access to  information theory  \cite{cover}.
In any case, the Shannon information entropy and quantum information
entropy ($IE$) are closely related. The first uses a normalized set of
probabilities, the other uses the density
matrix formalism  and a normalized trace. The density matrix is
diagonalized and the eigenvalues are
taken as probabilities to calculate $IE$
\begin{equation}
S=-\sum_{i=1}^n p_i \ln  p_i,\quad \mbox{with}\quad \sum_{i=1}^n
p_i=1.
\end{equation}

The eigenvectors are generally not used to define a probability
distribution, but investigations along such line have been done
and produced interesting results and conclusions. To distinguish
their input approach, from the full density matrix $IE$, it is
said, Shannon information entropy is  calculated, with entities in
{\em r-} and {\em k-}space, for atoms \cite{gadre},  and more
recently for nuclei, atomic clusters and boson traps \cite{massen}
in particular.

In subnano physics $IE$ is everywhere present.
It reduces radically any genuine
and detailed information of a many particle system and
the normalization condition eliminates any aspect of  relative
scales which are existing within physical quantities and are the key
for prevalence and subordinate. Within the set $p_i$, the succession is
irrelevant.
As such,  the pure mathematical aspect of $IE$ is impossible to escape and
any set of positive definite normalized physical data
are used as a potential set $\{p_i\}$. Thus, the resulting information
entropy does not automatically have  a  physical interpretation
since an abstract
scheme of order or disorder is now in vogue.

With this outline in mind, we  do not hesitate to
augment nucleon-nucleus $(NA)$ cross sections with the
expression of information entropy. In this way we concatenate
geometry, kinematics and dynamics of nuclear reactions. We expect that
these entities are separable and thus imply a linear
relation for measures of $IE$.

This conjecture is based upon a uniform probability distribution, to simplify
the $IE$ relation,  for all
partial wave scattering amplitudes, which form our input data
$p_i=\frac 1 n $ and $i=1,\dots,n$, where
$ n-1=l_{max}=k R=k x_0 A^{1/3} $
is the grazing (largest relevant)
angular momentum. This is determined from the grazing radius
$R=x_0 A^{1/3}$, for which a nuclear Fermi distribution implies
$x_0=r_0+3a/A^{1/3}$ with $r_0=1.15$\,[fm] and $a=0.55$\,[fm].
The  masses (projectile, target) $m_1$, $m_2$ and $T=T_{lab}$
determine the wave number
\begin{eqnarray}
(\hbar c)^2 k^2 &= & \frac{m_2^2 (T^2 + 2 m_1 T)}{ (m_1+m_2)^2 +2 m_2
 T} \nonumber\\   &=&  \frac{2m_1 m_2^2}{(m_1 +m_2)^2 } T +
\frac{(m_1 -m_2)^2 m_2^2}{(m_1+m_2)^4} T^2 + \cdots \nonumber\\
&\approx & \frac{2m_1 m_2^2}{(m_1 +m_2)^2 } T .
\end{eqnarray}

A uniform distribution, $\{ p_i=1/n\}$, implies
$IE$ to be
$S=-\sum_{i=1}^n \frac 1 n \ln \frac 1 n= \ln n$.
Together with the  approximated $k$ and $R$ values, this gives a desirable
linear dependence
\begin{equation}
S = a + \frac 1 2 \ln T + \frac 1 3 \ln A
\label{simple-ie}
\end{equation}
with
\begin{equation}
 a = \ln \frac{x_0 \sqrt{2 m_1}}{\hbar c} +
\ln \frac{m_2}{(m_1+m_2)}  .
\end{equation}

Eq. (\ref{simple-ie}) is interesting in itself, as it recalls the
well known physiological sensitivity of biological sense-organs
which often show a logarithmic dependence on magnitudes, i.e.
sound  pressure or light intensity. As one knows, animals alike
humans are very able to notice and detect small structural
differences, even within a chaotic environment, thus we anticipate
propitious support also from information entropy for the physics
of many body systems and in particular for quantum scattering. It
is not difficult to improve Eq. (\ref{simple-ie}) and foresee
departures from simplicity with nonlinear  dependencies between
the variables ($m_1,m_2,T,N+Z,N-Z$). To this end, we performed
accurate numerical studies with realistic input data and compiled
results for many nuclei and projectile  energies, ranging from low
to medium energies, $T_{lab}<1$\,[GeV].

The following sections contain the relevant scattering theory,
related  input data, numerical results and the final summary and
conclusions.

\section{Link of $NA$ scattering with $IE$}

The $NA$ scattering
amplitudes for zero spin targets
\begin{eqnarray}
f(\theta)&=& \frac{i}{2k}\sum_{l=0}^{lmax}
\big[(l+1)(1-\eta_l^+)+l(1-\eta_l^-)\big]
e^{i\sigma_l}P_l(\cos \theta) \nonumber\\
  & &-\frac1{2k}\sum_{l=0}^{lmax}
(\eta_l^+-\eta_l^-)e^{i\sigma_l}\frac{d}{d\theta}P_l(\cos \theta)+f_c(\theta),
\label{scatt-ampl}
\end{eqnarray}
are readily available for any type of cross section and
spin observable  \cite{amos,funk}. In this application
we suppress the  Coulomb amplitude
$f_c(\theta)$ and use only  the sum of
partial wave amplitudes. The associated angle integrated
cross sections, $(e)$ for
elastic, $(r)$ for reaction and $(t)$ for total, are
\begin{equation}
\sigma^{(t)}=\sigma^{(e)} +\sigma^{(r)}= \sum_{l=0}^{lmax}
\sigma^{(e)}_{l} + \sum_{l=0}^{lmax}\sigma^{(r)}_{l},
\end{equation}
where partial wave cross sections are
 \begin{eqnarray}
\sigma_{l}^{(e)} &=& \frac{\pi}{k^2}
\big[ (l+1)|1-\eta_l^+|^2+l|1-\eta_l^-|^2 \big],\\
\sigma^{(r)}_{l}&=&
\frac{\pi}{k^2}\big[(l+1)(1-|\eta_l^+|^2)+l(1-|\eta_l^-|^2) \big].
\end{eqnarray}
After normalization, by $\sigma^{(t)}$, the partial wave
cross section probabilities distinguish contributions from
elastic and reaction channels
\begin{equation}
1=\sum_{l=0}^{lmax} \big[p_l^{(e)} + p_l^{(r)}\big],
\end{equation}
and
\begin{equation}
 p_l^{(e)}=\frac{\sigma_l^{(e)}}{\sigma^{(t)}}, \quad \mbox{and}\quad
 p_l^{(r)}=\frac{\sigma_l^{(r)}}{\sigma^{(t)}}.
 \end{equation}
The $l-$dependent probabilities $p_l^{(e)}$ and $p_l^{(r)}$ include no
interference between different $l-$values, the density matrix is diagonal
and off-diagonal elements are zero. Thus
the information entropy for  proton and neutron nucleus scattering
integrated cross sections are a straight forward sum of
contributing angular momenta $0\le l \le l_{max}$
\begin{equation}
S= - \sum_{l=0}^{lmax} \big[p_l^{(e)} \ln p_l^{(e)} +p_l^{(r)} \ln
  p_l^{(r)} \big].
\label{ie-eqn}
\end{equation}

An ample remark. In case of angle and complex spin dependent scattering
and reactions it is necessary to use the full power of the density
matrix formalism with diagonalization.
This formalism is well developed and general
scattering programs,
for low and medium energy scattering, are available. A range of such
programs has been
developed by J. Raynal, CEN-Saclay, with older as well as more current versions of DWBA,
with  a sophisticated microscopic approach  suited for
$NA$ scattering, and  versions of the  coupled channels
code ECIS  \cite{raynal}.

For introductory studies, we suggest to use the scattering amplitude
given by Eq. (\ref{scatt-ampl}). These amplitudes can easily be arranged in a
matrix
\begin{equation}
\rho(l_1,j_1,l_2,j_2|T,\theta)= \frac{|l_1,j_1,T,\theta \rangle
\langle l_2,j_2,T,\theta|} {\rm{Tr\  } \rho(l,j,l,j|T,\theta)}
\end{equation}
and  diagonalized.

\section{Application}

There exist many theoretical and  experimental
studies of  $NA$  scattering
amplitudes. Most of them are calculated with a Schr\"odinger
equation and optical model potential
fitted to data. Today, the potentials are
microscopic optical potentials. These optical potentials are based upon
a high quality $NN$ potential. Such a microscopic approach is our choice for this
study and  we are using the Argonne AV18 $NN$ potential  \cite{av18}.
Hereby, we generalized AV18 above pion
production threshold to become a complex $NN$ optical potential \cite{funk}.
Globally, these $IE$ studies are insensitive to the use of a {\it t-}
or {\it g-}matrix  microscopic  optical model potential \cite{amos,arellano}.
Undoubtedly, phenomenological optical model potentials are
bound to yield results which confirm ours.

The herein used
scattering amplitudes contain data for
$(n,A)$ and $(p,A)$ scattering on a dense grid of energies, $ 20 <
T(n,p)< 1000$ [MeV],  and
targets,  $^{4}$He, $^{12}$C, $^{16}$O, $^{40}$Ca, $^{58}$Ni,
$^{90}$Zr and $^{208}$Pb. The Coulomb amplitude $f_c$ is suppressed
in case of $(p,A)$.  The results are shown
in Fig.\,1, with the calculated $IE$ values (circles), for our selection
of nuclei and energies.  We verified, with the numerical $IE$ results,  a linear
dependence, as given in Eq. (\ref{simple-ie}),  and
determined, with a  $\chi^2$ fit, the expansion coefficients

\begin{equation}
\chi^2 = \rm{Min}_{(a,b,c)} \|S - (a+b\ln T+c\ln A) \|.
\label{iefit-eqn}
\end{equation}

The parameters $a$, $b$ and $c$ are given, for two energy fit
ranges, in Table\,I and Fig.\,1  shows lines of the linear
approximation together with the numerical results for the
representative set of nuclei and $T_{lab}$ energies.

\begin{table}
\caption{Eqs. (\ref{simple-ie}) and (\ref{iefit-eqn}),  linear relation
  best fit parameters.}
\begin{tabular}{ccccc}
 \hline  \hline
 Reaction Type   &  $\quad$a$\quad$   & $\quad$b$\quad$
 &$\qquad$c$\qquad$
& Remarks   \\
    \hline
   nA   and   pA   & -1.25$\pm$0.25 & 1/2 & 1/3 & uniform model
\\[1ex]
  nA   &   -1.0796  & 0.7168& 0.3787&   20-1000  MeV\\
  pA   &   -1.1866  & 0.7412& 0.3634&     20-1000  MeV
\\[1ex]
  nA   &   -1.1267  & 0.7348& 0.3598&   50-500  MeV\\
  pA   &   -1.2638  & 0.7638& 0.3485&     50-500  MeV  \\
  \hline
    \hline
\end{tabular}
\vspace*{-5mm}
\end{table}

\begin{widetext}

\begin{figure}[h]
\vspace*{-1.6cm}
\begin{center}
\begin{tabular}{cc}
  \includegraphics[width=6cm]{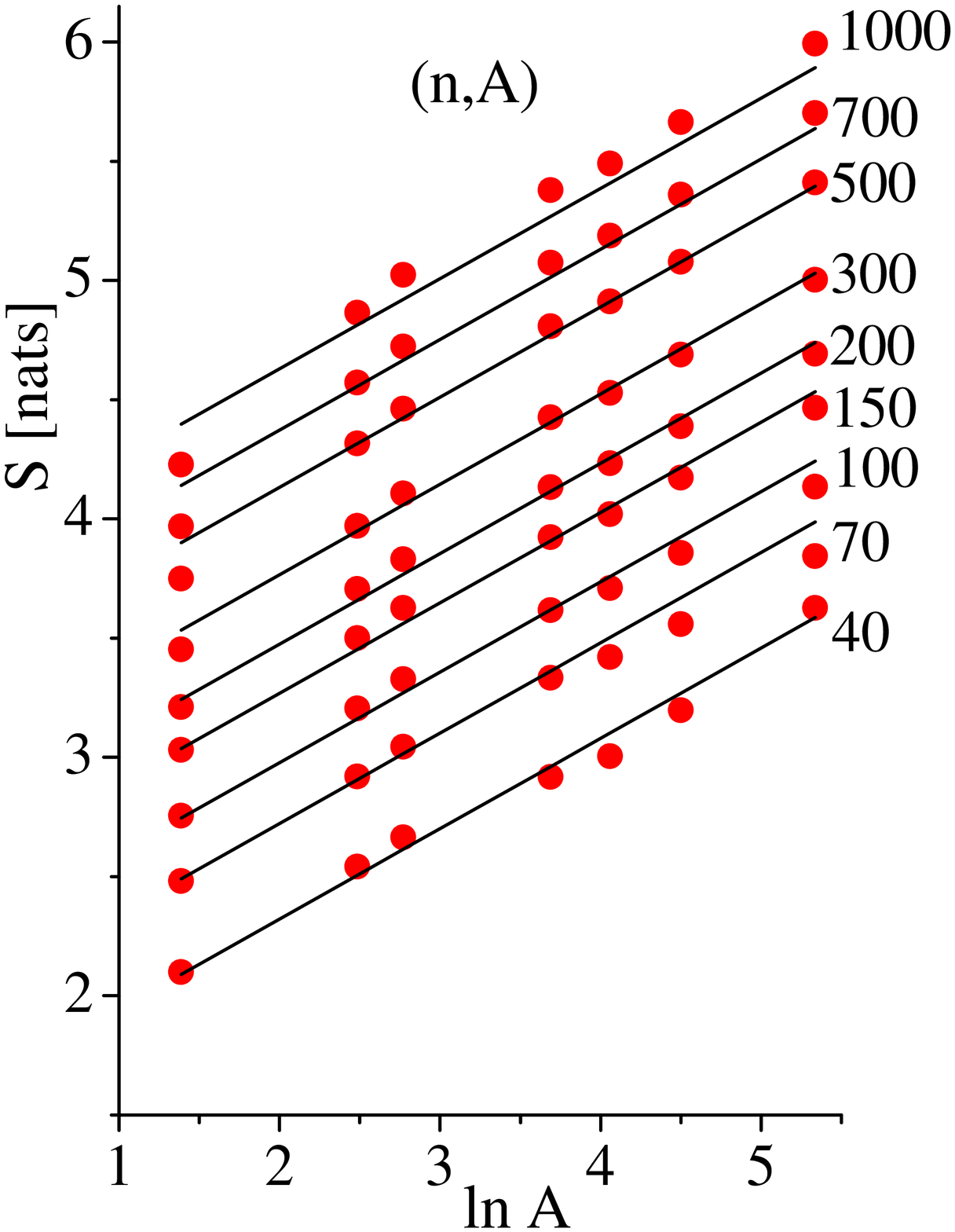}
&   \includegraphics[width=6cm]{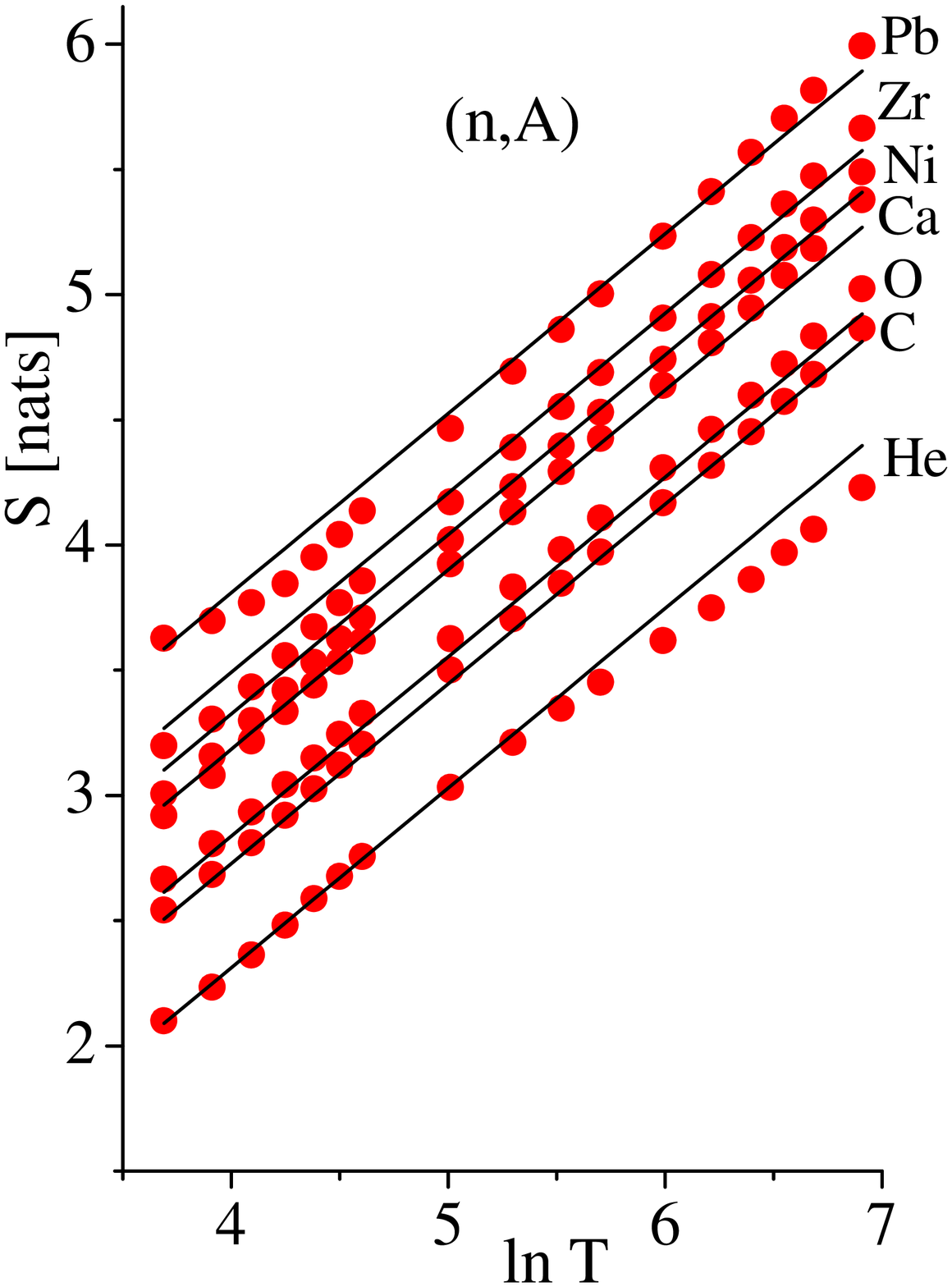}\vspace*{-15mm}\\
\includegraphics[width=6cm]{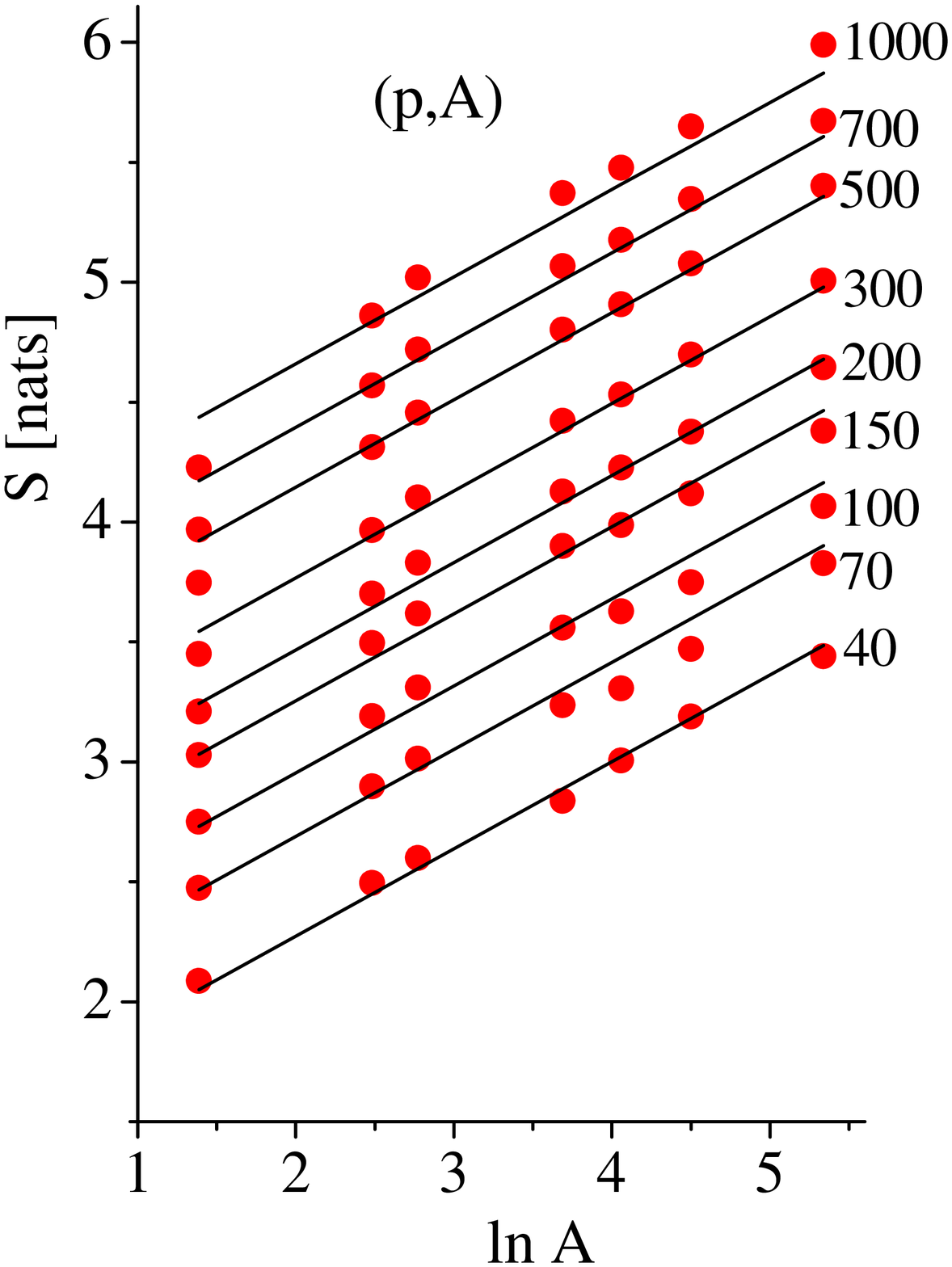}
&
\includegraphics[width=6cm]{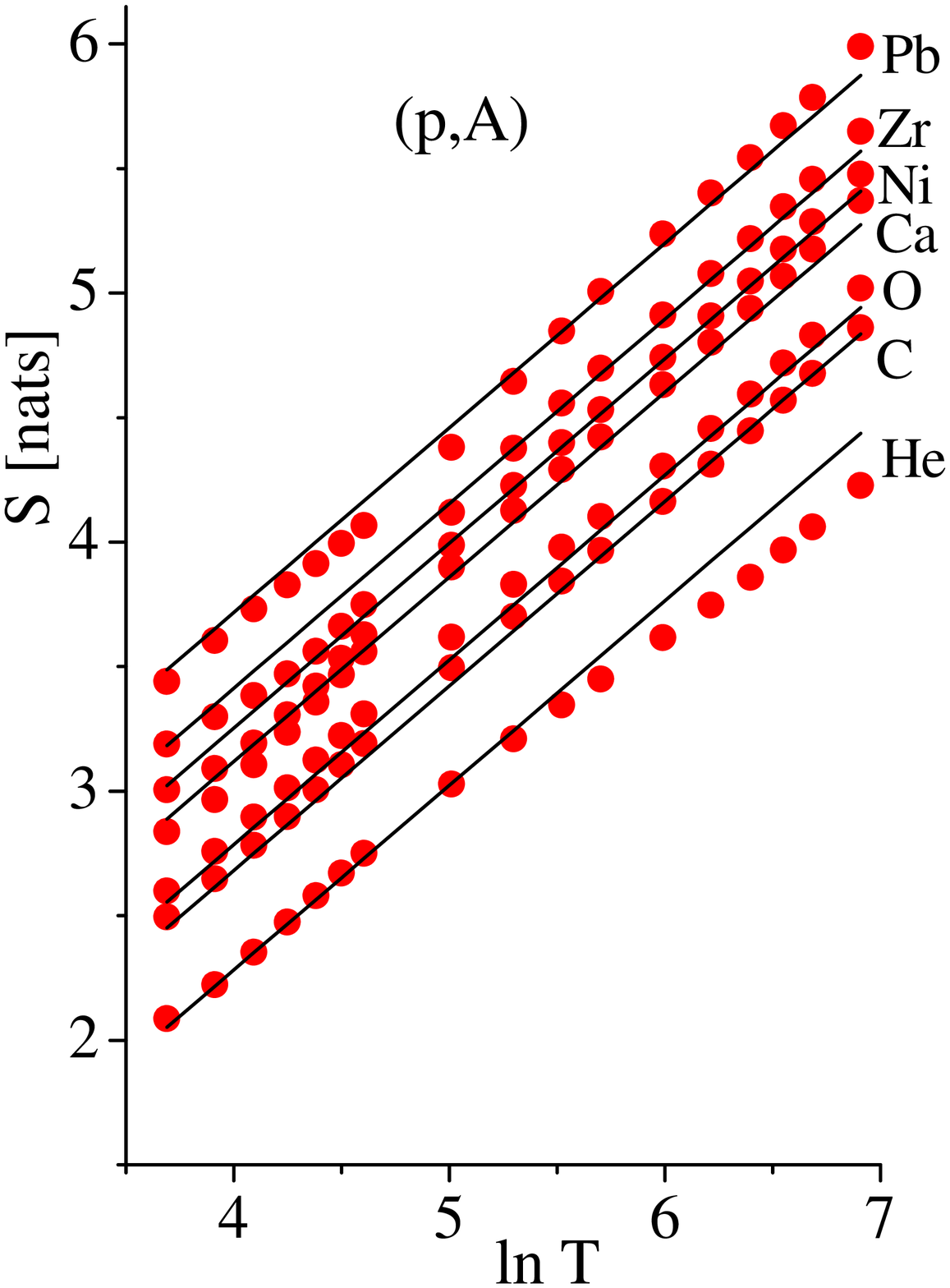}
\end{tabular}

\vspace*{-7mm}
 \caption{Verification of a linear dependence, using parameters associated with
the range $20-1000$\,[MeV] in Table\,I, versus
 $\ln T$ and $\ln A$, of
numerically calculated information entropy (red dots).}
 \label{fig-scat-1}
\end{center}
\vspace*{-0.5cm}
\end{figure}

\end{widetext}

It is possible to see some data scattering
around the lines. This is caused  by the  not smooth but
realistic dependence of microscopic optical model potentials  on
$A$ and $N-Z$ and the spectroscopy of the target states.
These particularities generate deviations from the lines  of a few
percent. Demanding  a better agreement would be unrealistic in the first place.

In summary, the analysis confirms a linear dependence of $IE$ in $NA$
scattering from  two entities,
first the {\em logarithm of
target A}  and second the {\em logarithm of projectile kinetic energy T}.
In particular, we suggest to use  the relation for $IE$
\begin{equation}
S(A,T)=a+b \ln T +c \ln A
\end{equation}
with best fit parameters $(a,b,c)$ from Table\,I.

Finally,  we present a  numerical
study and comparison of $IE$ for various isotopes, $S(A,N-Z,T)$. Such study is
bound to reflect the spectroscopy of a  range of targets.
The numerical results are shown in Fig.\,2. Notice, in comparison
with Fig.\,1, we have increased our
resolution in Fig.\,2   by an order of magnitude and
differentiate between scattering
of neutrons and protons with full circles and squares, respectively, all having $T=400$ [MeV].
The results are smooth and reflect shell closures for cases when  the neutron
number equals a magic number, {\em i.e.}    O at N=8,  Ca at N=20 and
28,  Ni at N=28 and 50,  Zr at N=50,  Sn at N=50 and 82, and for
Pb at N=126.

\begin{widetext}

\begin{figure}[h]
\begin{center}
\begin{tabular}{ccc}
  \includegraphics[width=5.5cm]{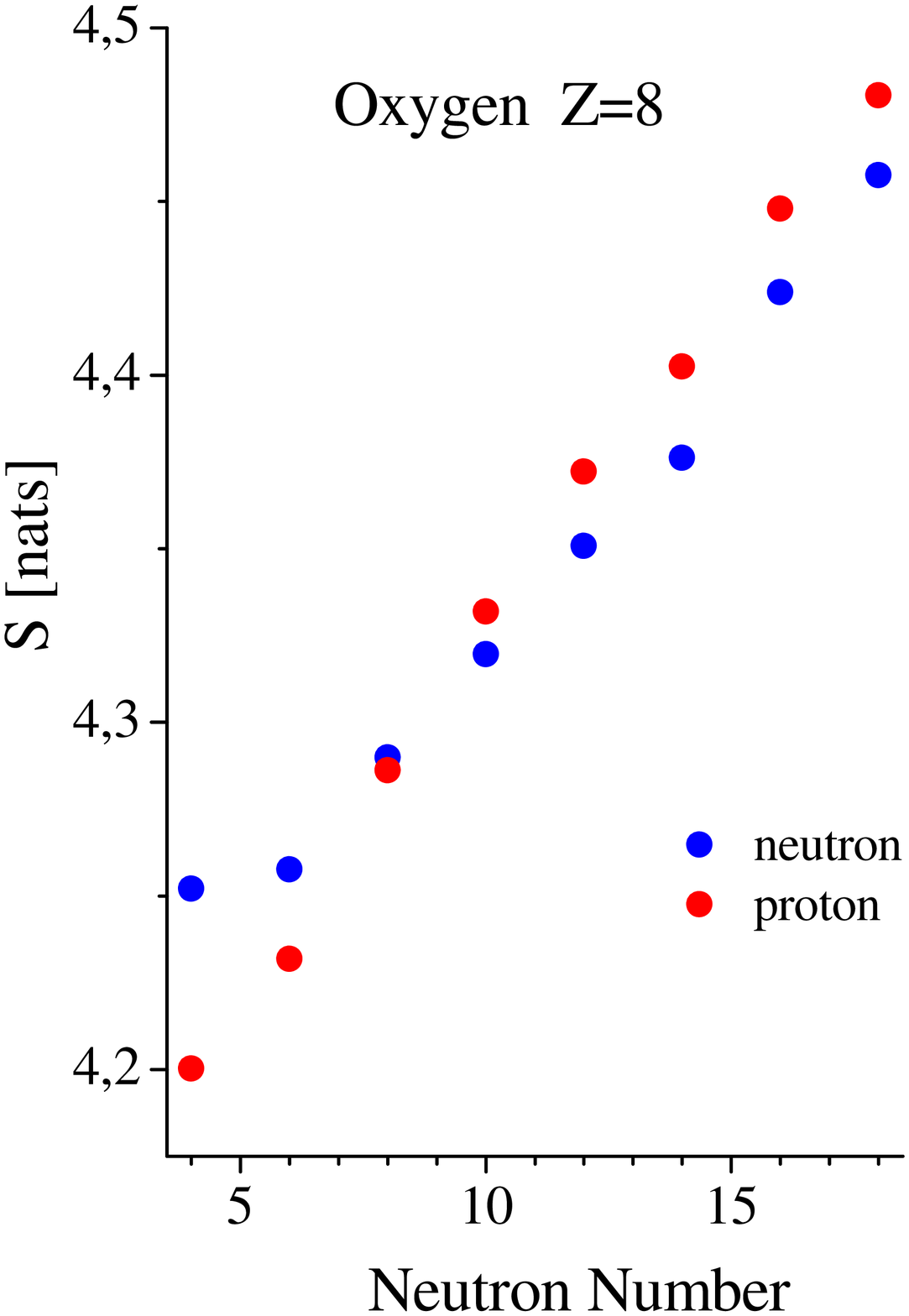}
&   \includegraphics[width=5.5cm]{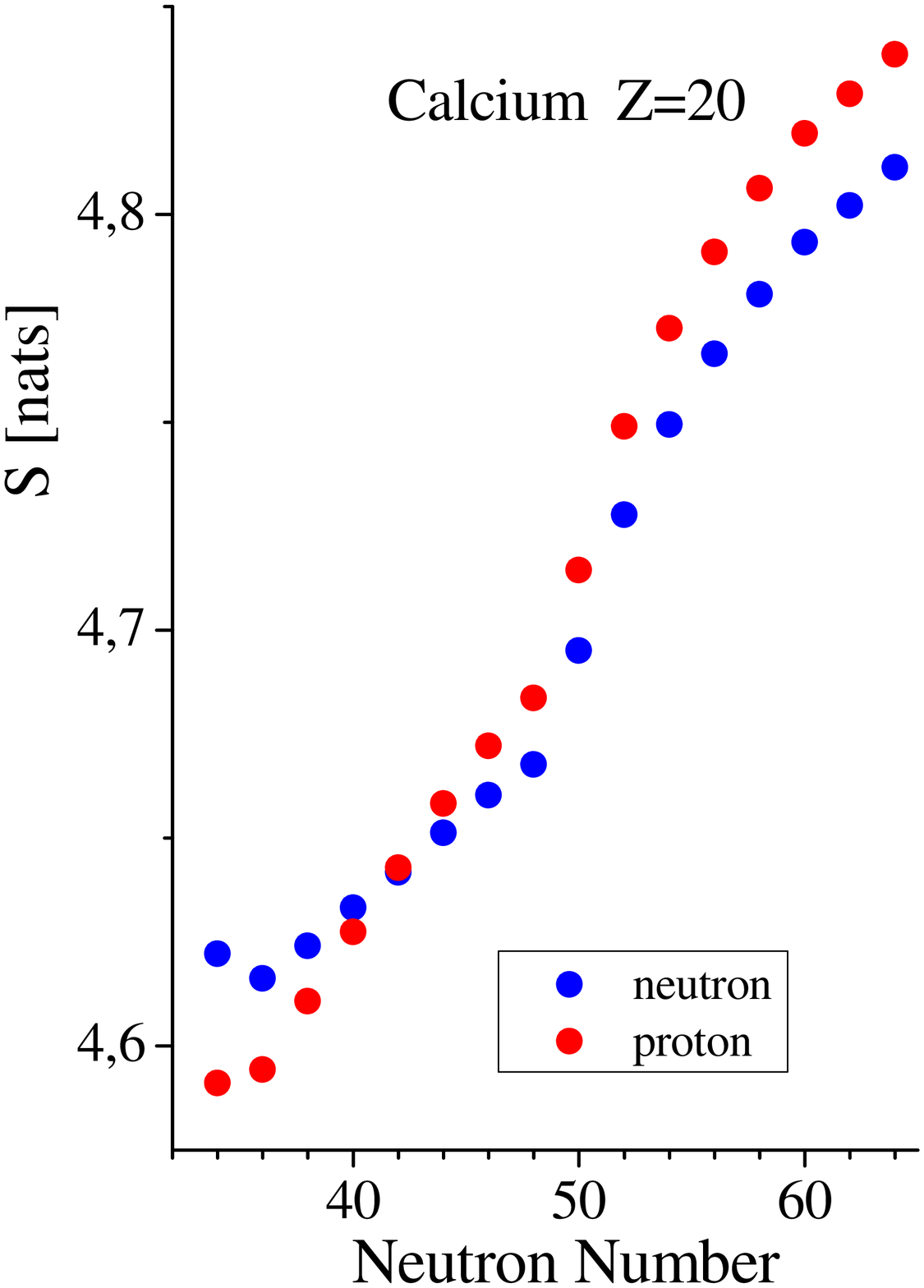} &
\includegraphics[width=5.5cm]{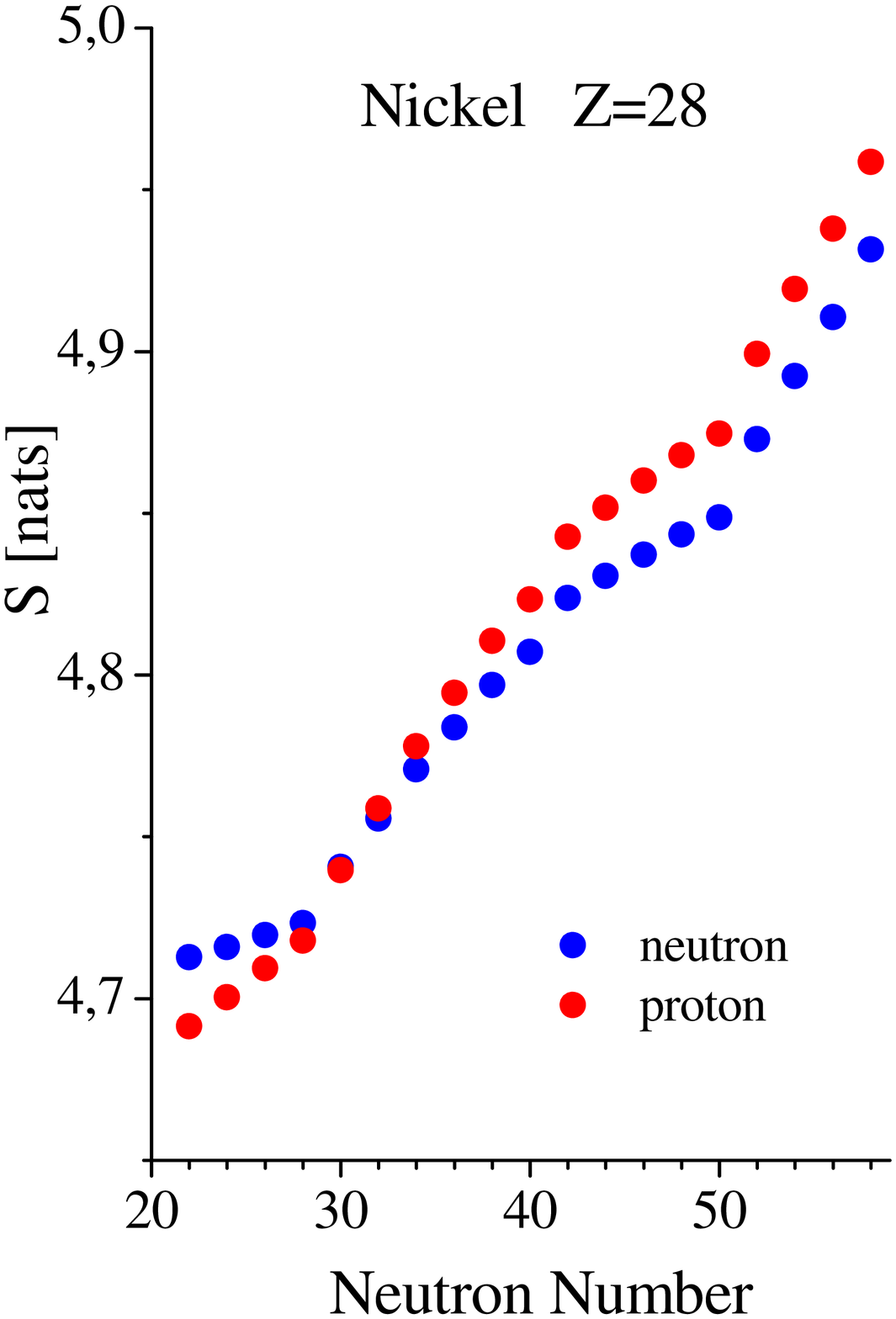}
\vspace*{-1cm}\\
  \includegraphics[width=5.5cm]{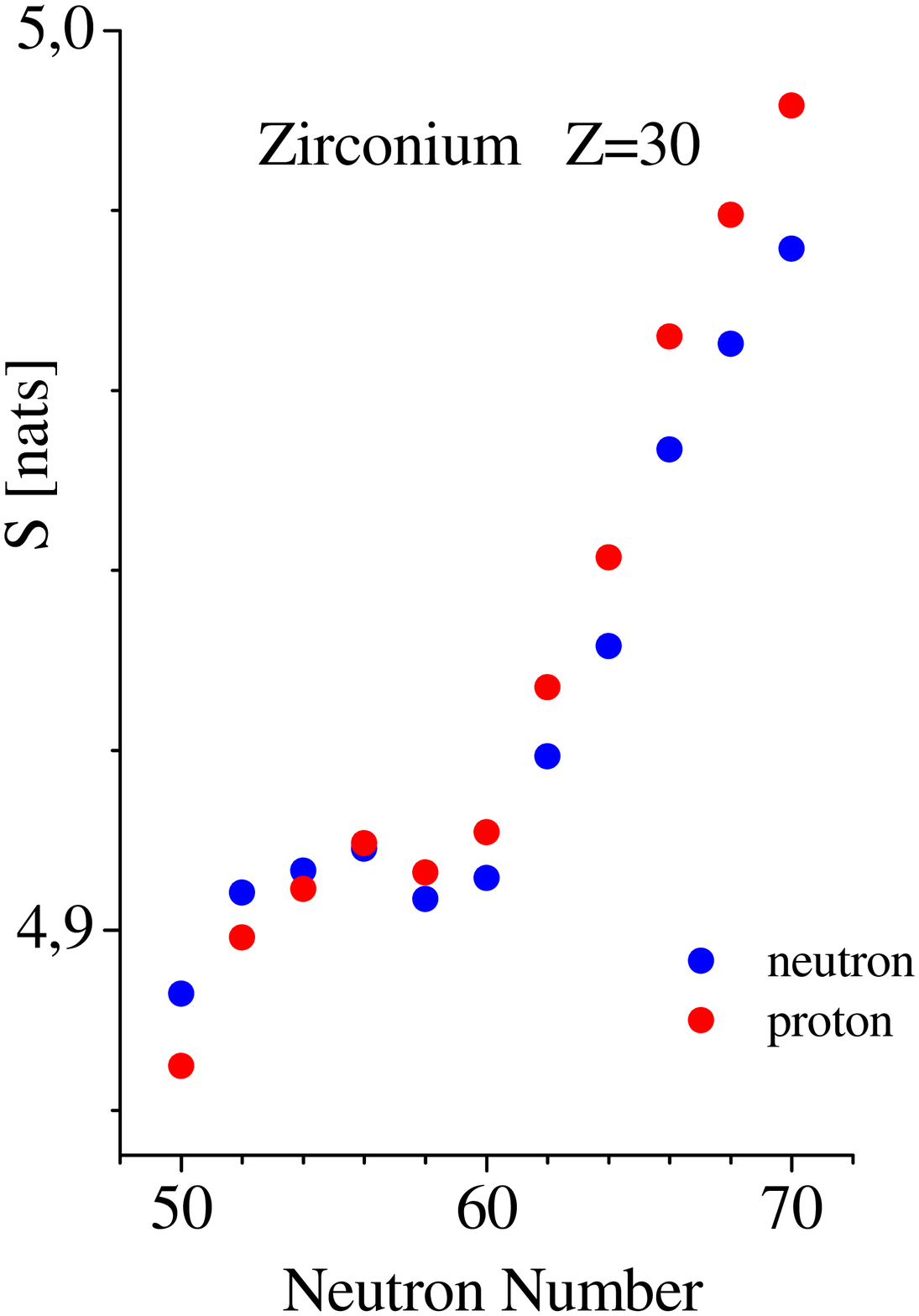}
&   \includegraphics[width=5.5cm]{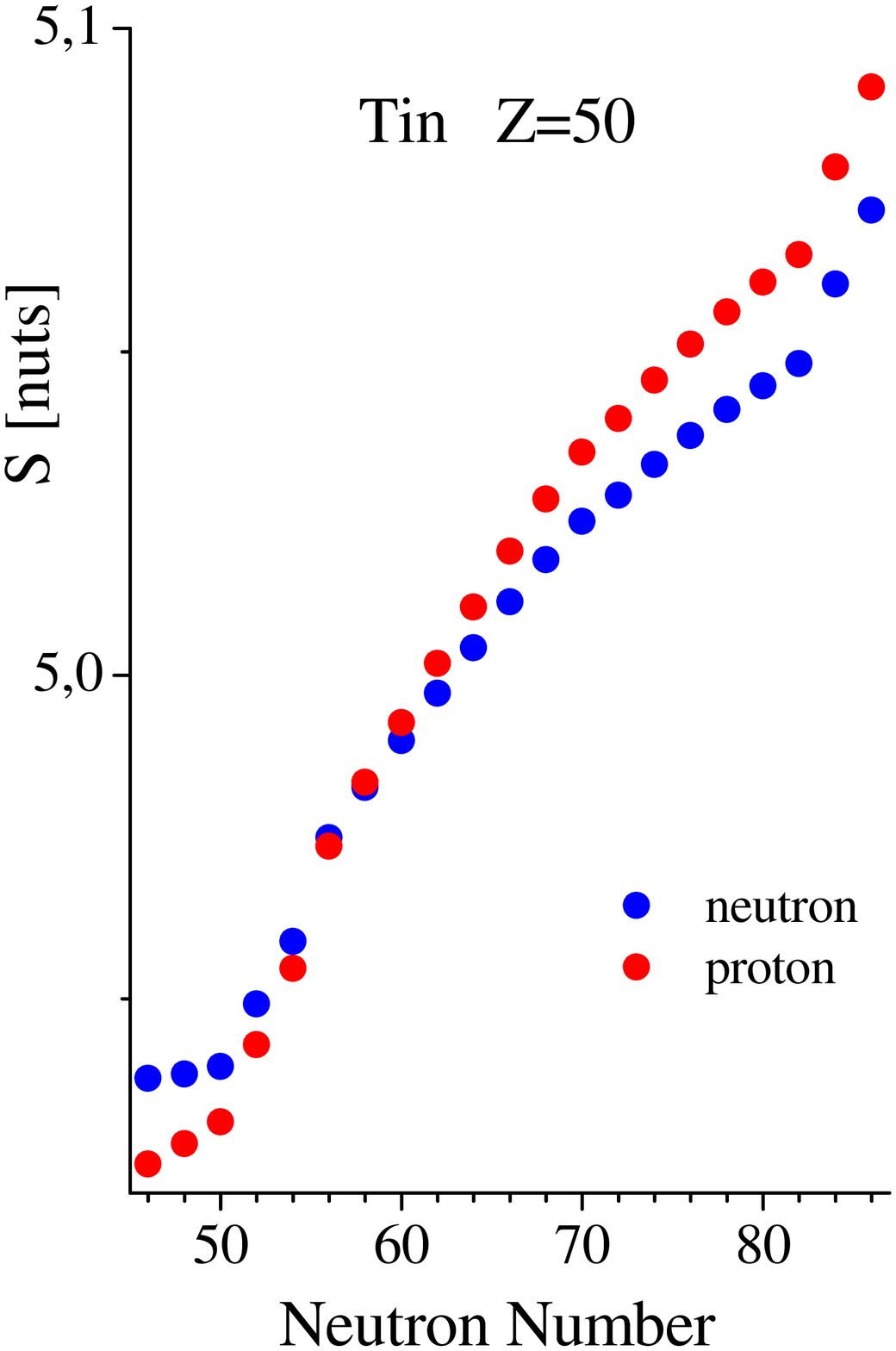} &
\includegraphics[width=5.5cm]{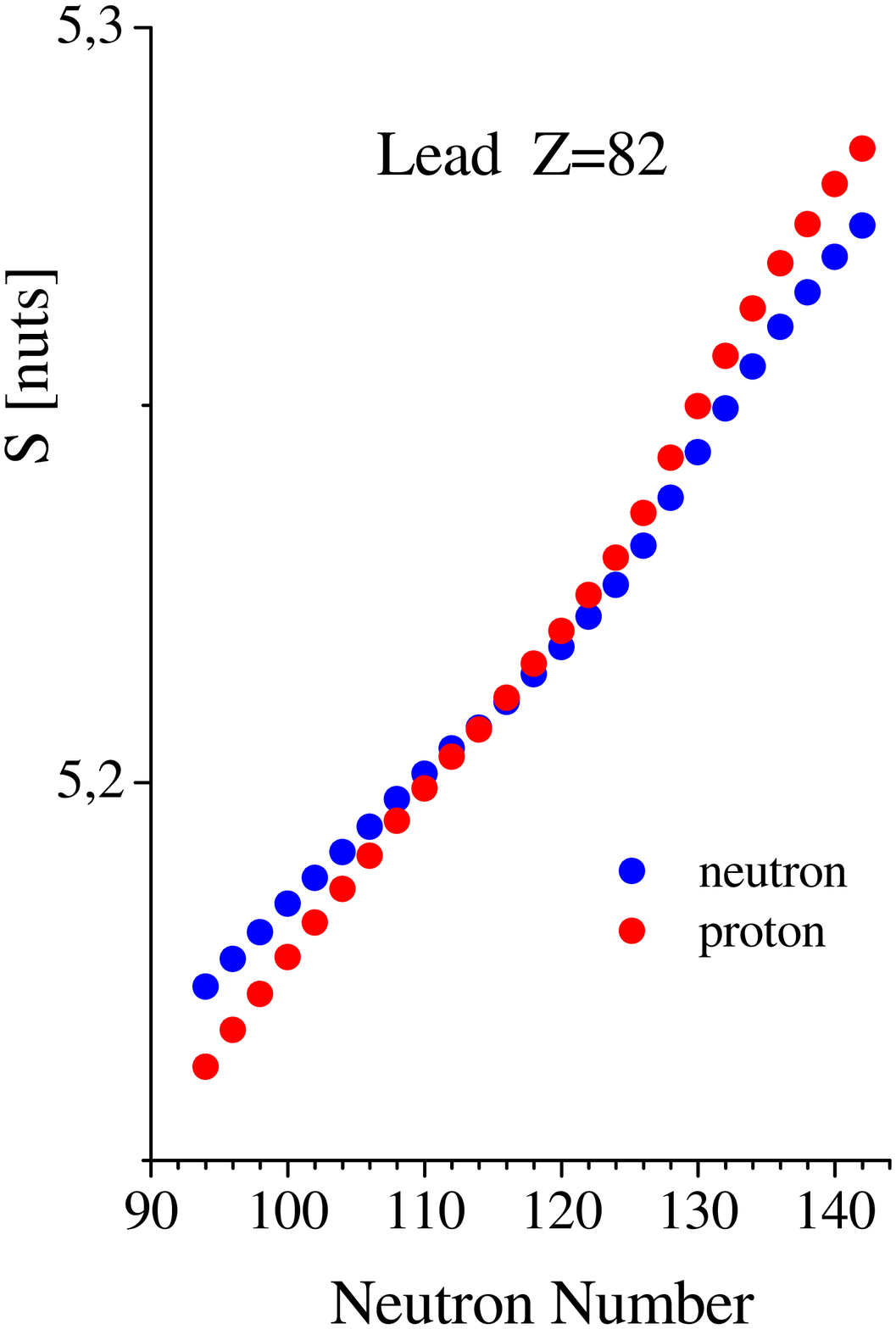}
\end{tabular}
\caption{Dependence of  information entropy $S(A,N-Z)$ for
O, Ca, Ni, Zr, Sn and Pb isotopes. Shown are  numerical results
for (n,A)  in (blue)  and (p,A) in (red), $T_{lab}=400$\,[MeV].}
 \label{fig-scat-2}
\end{center}
\end{figure}

\end{widetext}

\section{Summary and conclusions}

Information entropy traces its roots to the work of Shannon in the
first place, as a fundamental result of classical telecommunication theory,
since that time it has received  much attention in experiment and theory, for
classical as well as  quantum systems  \cite{nielsen,cover}.
$IE$, as a mean to quantify  entanglement
of quantum mechanical states, of bound or scattering states,
encouraged also our study with the aim to find smooth and simple
structures to emerge in the least structured quantity of
nucleon-nucleus scattering, elastic and reaction cross sections
at low to medium energies 20-1000 [MeV].

The results of this study predict a smooth qualitative dependence of $IE$ on
$ \ln T_{lab}$ and $\ln A$. The observed numerical scattering, around this
smooth and averaged results, are first due to shell effects
of the target and second due to target mass and charge
dependencies, i.e. functions of $(N+Z)$ and $(N-Z)$ \cite{arellano}.
The mathematical
nature of $IE$ implies uncountable many applications in theoretical
studies with powerful options for intuitions and conjectures beyond the
known scattering analysis.

\section*{Acknowledgment}

We acknowledge several discussions with Dr. C.P. Panos, and  HVG
with gratitude the hospitality he received during his stay in the
Department of Theoretical Physics, Aristotle University of
Thessaloniki.

\end{document}